\documentclass[12pt]{article}
\usepackage{cite}
\usepackage{CJK}
\usepackage{amssymb}
\usepackage{amsmath}
\usepackage{enumerate}
\usepackage{graphicx}
\usepackage{amsfonts}
\usepackage{url}
\usepackage{bm}

\usepackage{pifont}
\usepackage{epsfig,subfigure,dsfont,amsthm,amsbsy,mathrsfs,amscd}

\def\la{\langle}
\def\ra{\rangle}
\def\be{\begin{equation}}
\def\ee{\end{equation}}
\def\ba{\begin{array}}
\def\ea{\end{array}}
\input amssym.def

\newcommand\btd{\raise 2pt \hbox{$\hat\bigtriangledown$}\hskip 1.5pt}
\newcommand\bt{\raise 2pt \hbox{$\bigtriangledown$}\hskip 1.5pt}
\begin{document}
 \title{\large\bf Detection and measure of genuine tripartite entanglement with partial transposition and realignment of density matrices}
\author{Ming Li$^\dag$, Jing Wang$^\dag$, Shuqian Shen$^\dag$, Zhihua Chen$^\ddag$, and Shao-Ming Fei$^\sharp$$^\S$\\[10pt]
\footnotesize \small $^\dag$College of the Science, China University
of Petroleum,\\
\small Qingdao 266580, P. R. China\\
\footnotesize
\small $^\ddag$Department of Science, Zhijiang college, Zhejiang University of Technology, \\
\small Hangzhou, 310024, P. R. China\\
\small $^\sharp$School of Mathematical Sciences, Capital Normal University,\\
\small Beijing 100048, P. R. China\\
\small $^\S$Max-Planck-Institute for Mathematics in the Sciences,\\
\small Leipzig 04103, Germany\\}
\date{}

\maketitle

\centerline{$^\ast$ Correspondence to jwang@upc.edu.cn}
\bigskip

\begin{abstract}
It is challenging task to detect and measure genuine multipartite entanglement. We investigate the problem by considering the average based positive partial transposition(PPT) criterion and the realignment criterion. Sufficient conditions for detecting genuine tripartite entanglement are presented. We also derive lower bounds for the genuine tripartite entanglement concurrence with respect to the conditions. While the PPT criterion and the realignment criterion are powerful for detecting bipartite entanglement and for providing lower bounds of bipartite concurrences,
our results give an effective operational way to detect and measure
the genuine tripartite entanglement.
\end{abstract}
\bigskip

Quantum entanglement is recognized as a remarkable resource in the rapidly
expanding field of quantum information science, with various
applications \cite{nielsen}.
A multipartite quantum state
that is not separable with respect to any bi-partition is
said to be genuinely multipartite entangled(GME) \cite{guhnerev},
which is one of the
important type of entanglement, and offers significant
advantage in quantum tasks comparing with bipartite entanglement
\cite{mule1}. In particular, it is the basic ingredient in
measurement-based quantum computation \cite{mule2}, and is
beneficial in various quantum communication protocols,
including secret sharing \cite{mule4,hillery}, extreme spin squeezing \cite{srensen}, high sensitivity in some
general metrology tasks \cite{toth}, quantum computing using
cluster states \cite{rauss}, and multiparty quantum network  \cite{mule3}.
Although its significance, detecting and measuring such kind of entanglement turns out to be quite
difficult.
To certify GME, an abundance of linear and nonlinear entanglement witnesses \cite{12, huber,
vicente3, huber1, wu, sperling, 14, claude, horo}, generalized
concurrence for multi genuine entanglement \cite{ma1,
ma2,gaot1,gaot2}, and Bell-like inequalities \cite{bellgme}entanglement witnesses
were derived (see e.g. reviews \cite{guhnerev,8}) and a characterisation in terms of semi-definite programs
(SDP) was developed \cite{jungnitsch,10}.
Nevertheless, the problem remains far from being satisfactorily solved.

For bipartite systems, Peres in \cite{ppt}has presented a much stronger separability criterion,
which is called positive partial transpose (PPT) criterion. It says
that if $\rho_{AB}$ is separable, then the partial transposition
$\rho_{AB}^{T_B}$ with matrix elements defined as:
$(\rho_{AB}^{T_B})_{ij,kl}=\rho_{il,kj}$
is a density operator (i.e. has
nonnegative spectrum). It has
interpretation as a partial time reversal \cite{san98}.
There is yet another strong class of criteria based on linear
contractions on product states. They stem from the new criterion
discovered in \cite{rudolph, ChenQIC03} called computable cross norm
criterion or matrix realignment criterion(CCNR) which is operational
and independent on PPT test \cite{ppt}. In terms of matrix elements
it can be stated as follows: if the state $\rho_{AB}$ is separable
then the matrix ${\cal R}(\rho)$ with elements
${\cal{R}}(\rho)_{ij,kl}=\rho_{ik,jl}$ has trace norm not greater than
one, i.e. $||{\cal{R}}(\rho)||_{KF}\leq 1.$ Quite remarkably,
the realignment criterion has been found to be able to detect some
PPT entangled states \cite{rudolph, ChenQIC03} and to be useful for
construction of some nondecomposable maps. It also provides nice
lower bound on concurrence \cite{chenprl}. Further more, a necessary and sufficient criterion of the local unitary equivalence for general multipartite states based on matrix realignment has been presented in \cite{zhangtgpra}.

In this manuscript, we investigate the detection of GME for arbitrary tripartite quantum systems. We will derive an effective criterion based on PPT and CCNR. A lower bound for GME concurrence will be also obtained. We then compute examples to show the effectiveness of our results.

\medskip
\noindent{\bf Results}
\medskip

In the following, we present a criterion to detect GME for tripartite qudits systems by using the PPT and CCNR criteria. A lower bound for GME concurrence of tripartite quantum systems will be also obtained. We start with some definitions and notations.

Let $H_i^d$, $i=1,2,3$, denote $d$-dimensional Hilbert spaces. A tripartite state $\rho  \in H_1^d \otimes H_2^d \otimes H_3^d$ can be expressed as $\rho  = \sum {{p_\alpha }} \left| {{\psi _\alpha }} \right\rangle \left\langle {{\psi_\alpha }} \right|$, where $0<p_\alpha\leq 1$,
$\sum {{p_\alpha }}  = 1$, $\left| {{\psi _\alpha }} \right\rangle \in H_1^d \otimes H_2^d \otimes H_3^d$ are normalized pure states.
If all $\left| {{\psi _\alpha }} \right\rangle$ are biseparable, namely, either
$\left| {{\psi _\alpha }} \right\rangle = \left| {\varphi _\alpha ^1} \right\rangle  \otimes \left| {\varphi _\alpha ^{23}} \right\rangle $ or $\left| {{\psi _\beta }} \right\rangle  = \left| {\varphi _\beta ^2} \right\rangle  \otimes \left|{\varphi _\beta ^{13}} \right\rangle $ or $\left| {{\psi _\gamma }} \right\rangle  = \left| {\varphi _\gamma ^3} \right\rangle  \otimes \left|{\varphi _\gamma ^{12}} \right\rangle$,
where $\left| {\varphi_\gamma^i} \right\rangle$ and $\left| {\varphi_\gamma^{ij}} \right\rangle$ denote pure states in $H_i^d$ and $H_i^d \otimes H_j^d$ respectively,
then $\rho $ is said to be bipartite separable. Otherwise, $\rho $ is called genuine multipartite entangled.

Define that ${M}(\rho ) = \frac{1}{3}(\left\| {{\rho ^{{T_1}}}} \right\| + \left\| {{\rho ^{{T_2}}}} \right\| + \left\| {{\rho ^{{T_3}}}} \right\|),$ $N(\rho ) = \frac{1}{3}(\left\| {{R_{\left. 1 \right|23}}(\rho )} \right\| + \left\| {{R_{\left. 2 \right|13}}(\rho )} \right\| + \left\| {{R_{\left. 3 \right|12}}(\rho )} \right\|),$ where $T_i$s are the partial transposition over the $i$th subsystem, $i=1,2,3$ and $R_{\left. i \right|jk}$ stands for the bipartite realignment with respect to subsystem $i$ and subsystems $jk$, $i,j,k=1,2,3$. $\left\|\cdot \right\|$ denotes the trace norm of a matrix.

To derive GME criterion, we first obtain the following lemma.

\textbf{\textit{Lemma:}} \label{lema} Let $d=\min\{m,n\}$.
For a bipartite quantum state $\left| \varphi  \right\rangle  \in {H_A^m} \otimes {H_B^n},$ we have $\left\| {{{\left( {\left| \varphi  \right\rangle \left\langle \varphi  \right|} \right)}^{{T_A}}}} \right\| \le d$, and $\left\| R_{A|B}(\left| \varphi  \right\rangle \left\langle \varphi  \right|) \right\|\le d$.

\textbf{Proof.} By Schmidt decomposition, we set $\left| \varphi  \right\rangle  = \sum\limits_{i = 1}^d {\sqrt {{u_i}} \left| {ii} \right\rangle } $ with $\sum\limits_{i = 1}^d {{u_i}}  = 1, u_i\geq0.$ By the Cauchy-Schwarz inequality one computes
\be\label{equa}\left\| {{{\left( {\left| \varphi  \right\rangle \left\langle \varphi  \right|} \right)}^{{T_A}}}} \right\| =\left\| R_{A|B}(\left| \varphi  \right\rangle \left\langle \varphi  \right|) \right\|= {(\sum\limits_i {\sqrt {{u_i}} } )^2}\leq d{(\sum\limits_i {{u_i}} )^2}=d.\ee \hfill \rule{1ex}{1ex}

Then we are ready to show the theorems.

\textbf{Theorem 1:} Let $\rho  \in {H_{123}} = H_1^d \otimes H_2^d \otimes H_3^d$ be a tripartite qudits quantum state.  If $\rho $ is bipartite separable, then $\max \{M(\rho ), N(\rho )\} \le \frac{1 + 2d}{3}$ must hold. Or equivalently, if $\max \{M(\rho ), N(\rho )\} > \frac{1 + 2d}{3},$ then $\rho $ is GME.

See Methods for the proof of theorem 1.

The GME concurrence for tripartite quantum systems, which is defined as follows, is proved to be a well defined measure\cite{ma1,ma2}.
For a pure state $|\psi\ra\in H_1^d \otimes H_2^d \otimes H_3^d$, the GME concurrence is defined by
\begin{eqnarray*}
C_{GME}(|\psi\ra)=\sqrt{\min\{1-tr(\rho_1^2),1-tr(\rho_2^2),1-tr(\rho_3^2)\}},
\end{eqnarray*}
where $\rho_i$ is the reduced matrix for the $i$th subsystem.
For mixed state $\rho\in H_1^d \otimes H_2^d \otimes H_3^d$, the GME concurrence is then defined by the convex roof
\begin{eqnarray}
C_{GME}(\rho)=\min\sum_{\{p_{\alpha},|\psi_{\alpha}\ra\}}p_{\alpha}C_{GME}(|\psi_{\alpha}\ra).
\end{eqnarray}
The minimum is taken over all pure ensemble decompositions of $\rho$.
Since one has to find the optimal ensemble to do the minimization, the GME concurrence is hard to compute.
In the following we derive an effective lower bound for GME concurrence in terms of the PPT criterion and the CCNR criterion.

\textbf{Theorem 2:}
Let $\rho  \in {H_{123}} = H_1^d \otimes H_2^d \otimes H_3^d$ be a tripartite qudits quantum state.
Then one has
\be
C_{GME}(\rho)\geq \frac{1}{\sqrt{d(d-1)}}(\max \{M(\rho ), N(\rho )\}-\frac{1+2d}{3}).
\ee

See Methods for the proof of theorem 2.

\medskip
\noindent{\bf Applications}
\medskip

The following two examples show that the criterion and the lower bound of GME concurrence above are much effective for detecting and measuring GME in tripartite quantum systems.

\textit{\textbf{Example 1:}}
Consider quantum state $\rho  \in H_1^3 \otimes H_2^3 \otimes H_3^3,$ $\rho  = \frac{1-x}{27}I + x\left| \varphi  \right\rangle \left\langle \varphi  \right|,$ where $\left| \varphi  \right\rangle  = \frac{1}{{\sqrt 3 }}(\left| {000} \right\rangle  + \left| {111} \right\rangle  + \left| {222} \right\rangle )$ is the GHZ state. By Theorem 1 in \cite{vicente3} we can detect GME for $0.894427<x\leq 1$. Using the Theorem 1 in this manuscript, we compute $\max \{M(\rho ), N(\rho )\}=\frac{1}{9}(8+10x+|10x-1|)$. Thus GME is detected for $0.7<x\leq 1$.

\textit{\textbf{Example 2: }}\label{ex31} We consider the mixture of the GHZ state and W state in three-qubit quantum systems $\rho= \frac{1-x-y}{8}I + x|GHZ\rangle\langle GHZ|+y|W\rangle\langle W|,$ where $|GHZ\rangle= \frac{1}{\sqrt 2}(|000\rangle  + |111\rangle)$ and $|W\rangle= \frac{1}{\sqrt 3}(|001\rangle  + |010\rangle+|100\rangle)$. As shown in Figure 1, our criterion detect some GME(blue region) that can not be detected by Vicente criteria.
\begin{figure}[h]
\begin{center}
\resizebox{8cm}{!}{\includegraphics{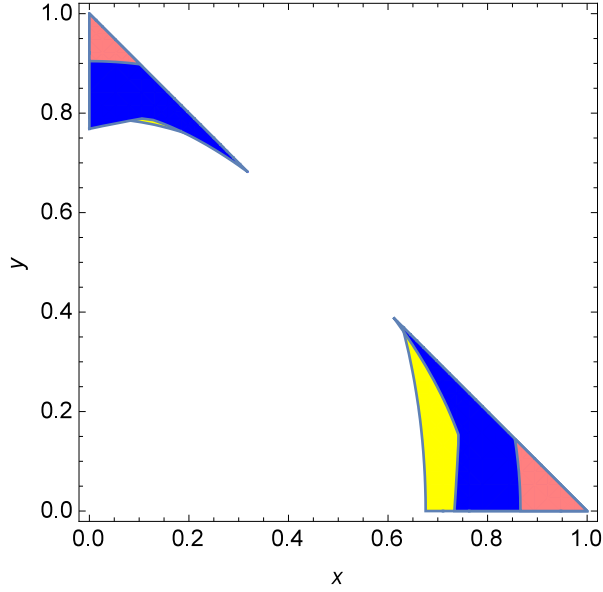}}
\end{center}
\caption{GME Detected by Vicente criterion (pink region by Theorem 1 and yellow region by Theorem 2 in \cite{vicente3}) and by the theorem 1 in this manuscript(blue region).\label{fig3}}
\end{figure}

The lower bound of GME concurrence in Theorem 2 for $\rho$ is computed to be
\begin{eqnarray*}
g(x,y)&=&(1/(24 \sqrt{2}))(-40 + 3 \sqrt{(-1 - 3 \alpha + \beta)^2} +
  6 \sqrt{(-1 + \alpha + \beta)^2}+  \sqrt{(3 - 3 \alpha + 13 \beta)^2}\\
&+&\sqrt{
  9 + 153 \alpha^2 + 6 \beta + 17 \beta^2 -
   6 \alpha (3 + \beta) -
   8 \sqrt{(3 - 3 \alpha + \beta)^2 (9 \alpha^2 + \beta^2)}}\\
&+&  \sqrt{9 + 153 \alpha^2 + 6 \beta + 17 \beta^2 -
   6 \alpha (3 + \beta) +
   8 \sqrt{(3 - 3 \alpha + \beta)^2 (9 \alpha^2 + \beta^2)}}\\
&+&\sqrt{9 + 45 \alpha^2 - 18 \alpha (-1 + \beta) - 18 \beta +
   137 \beta^2 -
   12 \sqrt{(1 + \alpha - \beta)^2 (9 \alpha^2 +
       32 \beta^2)}}\\
&+& \sqrt{
  9 + 45 \alpha^2 - 18 \alpha (-1 + \beta) - 18 \beta +
   137 \beta^2 +
   12 \sqrt{(1 + \alpha - \beta)^2 (9 \alpha^2 + 32 \beta^2)}})
\end{eqnarray*}
as ploted in Figure 2.
\begin{figure}[h]
\begin{center}
\resizebox{8cm}{!}{\includegraphics{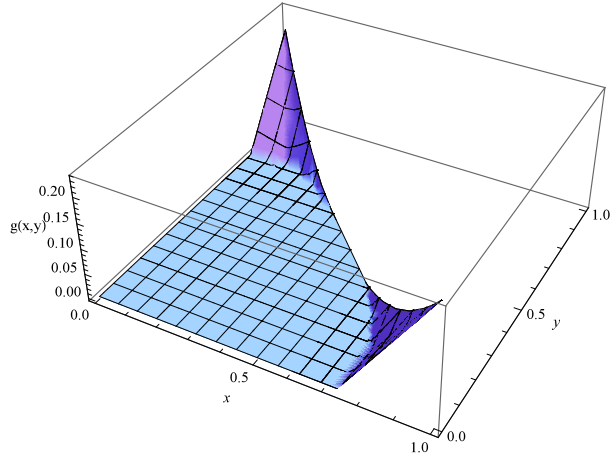}}
\end{center}
\caption{The lower bound of GME concurrence for $\rho$ in example 2. $g(x,y)$ stands for the lower bound.\label{fig3}}
\end{figure}

\medskip
\noindent{\bf Discussions}
It is a basic and fundamental question in quantum information theory to detect and measure GME. In this manuscript we have presented a GME criterion based on the PPT and Realignment criteria. A lower bound of GME concurrence for tripartite quantum system has also been obtained. Examples show that our criterion is independent of Vicente criteria and can detect more genuine entangled quantum states. Our results are derived by average based PPT and CCNR criteria. One can construct more effective criteria to detect GME and lower bounds of GME concurrence by taking the average of the correlation matrices or  covariance matrices and so on. It is also of interesting to investigate the implementation of the criterion with measurements or to extend the results to systems consisting of more than three parties.

\medskip
\noindent{\bf Methods}
\medskip

{\sf Proof of Theorem 1}~
 Lets consider a pure state $\rho=\left| \psi  \right\rangle \left\langle \psi  \right|$ first. Assume that $|\psi\ra\in {H_{123}} = H_1^d \otimes H_2^d \otimes H_3^d$ be bipartite separable, which will be in one of the following three forms: $| \psi  \rangle  = |\varphi_{1}\ra\otimes|\varphi_{23}\ra, |\psi\rangle  =|\varphi_{2}\ra\otimes|\varphi_{13}\ra$, or $\left| \psi  \right\rangle  = |\varphi_{3}\ra\otimes|\varphi_{12}\ra$.
If $\left| \psi  \right\rangle  = |\varphi_{1}\ra\otimes|\varphi_{23}\ra,$ then by using the first two equations in (\ref{equa}) we have
\begin{eqnarray*}
\left\| {{\rho ^{{T_1}}}} \right\| =\left\| R_{1|23}(\rho) \right\|= \left\| {{{\left( {\left| {{\varphi _1}} \right\rangle \left\langle {{\varphi _1}} \right|} \right)}^{{T_1}}} \otimes \left| {{\varphi _{23}}} \right\rangle \left\langle {{\varphi _{23}}} \right|} \right\|= 1;
\end{eqnarray*}
\begin{eqnarray*}
\left\| {{\rho ^{{T_2}}}} \right\| =\left\| R_{2|13}(\rho) \right\|= \left\| {\left| {{\varphi _1}} \right\rangle \left\langle {{\varphi _1}} \right|} \right\| \cdot \left\| {{{\left( {\left| {{\varphi _{23}}} \right\rangle \left\langle {{\varphi _{23}}} \right|} \right)}^{{T_2}}}} \right\| = \left\| {{{\left( {\left| {{\varphi _{23}}} \right\rangle \left\langle {{\varphi _{23}}} \right|} \right)}^{{T_2}}}} \right\| \le d;
\end{eqnarray*}
\begin{eqnarray*}
\left\| {{\rho ^{{T_3}}}} \right\|=\left\| R_{3|12}(\rho) \right\| = \left\| {\left| {{\varphi _1}} \right\rangle \left\langle {{\varphi _1}} \right|} \right\| \cdot \left\| {{{\left( {\left| {{\varphi _{23}}} \right\rangle \left\langle {{\varphi _{23}}} \right|} \right)}^{{T_2}}}} \right\|\le d.
\end{eqnarray*}
Similarly, one has
\begin{eqnarray*}
\left\| {{\rho ^{{T_1}}}} \right\| =\left\| R_{1|23}(\rho) \right\|\le d; \left\| {{\rho ^{{T_2}}}} \right\| =\left\| R_{2|13}(\rho) \right\|=1; \left\| {{\rho ^{{T_3}}}} \right\|=\left\| R_{3|12}(\rho) \right\|\le d
\end{eqnarray*}
for $\left| \psi  \right\rangle  = \left| {{\varphi _2}} \right\rangle \left\langle {{\varphi _{13}}} \right|$ and
\begin{eqnarray*}
\left\| {{\rho ^{{T_1}}}} \right\| =\left\| R_{1|23}(\rho) \right\|\le d; \left\| {{\rho ^{{T_2}}}} \right\| =\left\| R_{2|13}(\rho) \right\|\le d; \left\| {{\rho ^{{T_3}}}} \right\|=\left\| R_{3|12}(\rho) \right\|=1
\end{eqnarray*}
for $\left| \psi  \right\rangle  = \left| {{\varphi _3}} \right\rangle \left\langle {{\varphi _{12}}} \right|$ respectively.
Thus for any type bipartite separable pure quantum state, we always have $M(\rho) \le \frac{{1 + 2d}}{3},$ and $N(\rho) \le \frac{{1 + 2d}}{3}.$

For mixed bipartite separable state $\rho$, by using the convex property of $M(\rho)$ and $N(\rho)$ we obtain
\be M(\rho )\leq\sum {{p_\alpha }M(\left| {{\psi _\alpha }} \right\rangle \left\langle {{\psi _\alpha }} \right|)}  \le \frac{{1 + 2d}}{3},\ee
and \be N(\rho )\leq \sum {{p_\alpha }N(\left| {{\psi _\alpha }} \right\rangle \left\langle {{\psi _\alpha }} \right|)}  \le \frac{{1 + 2d}}{3},\ee
which proves the theorem.\hfill \rule{1ex}{1ex}

{\sf Proof of Theorem 2}~

Still we consider a pure state first.
Let $\rho=|\psi\ra\la\psi|\in H_1^d \otimes H_2^d \otimes H_3^d$ be a pure quantum state.
From the result in \cite{chenprl}, we have
\be
\sqrt{1-tr\rho_1^2}\geq  \frac{1}{\sqrt{d(d-1)}}(||\rho^{T_1}||-1);
\ee
\be
\sqrt{1-tr\rho_2^2}\geq  \frac{1}{\sqrt{d(d-1)}}(||\rho^{T_2}||-1);
\ee
\be
\sqrt{1-tr\rho_3^2}\geq  \frac{1}{\sqrt{d(d-1)}}(||\rho^{T_3}||-1).
\ee
One computes
\begin{eqnarray*}
&&3\sqrt{d(d-1)}\sqrt{1-tr\rho_1^2}-3\max \{M(\rho ), N(\rho )\}+1+2d\\
=&&3\sqrt{d(d-1)}\sqrt{1-tr\rho_1^2}-(||\rho^{T_1}||+||\rho^{T_2}||+||\rho^{T_3}||)+1+2d\\
\geq&&3\sqrt{d(d-1)}\sqrt{1-tr\rho_1^2}-\sqrt{d(d-1)}(\sqrt{1-tr\rho_1^2}+\sqrt{1-tr\rho_2^2}+\sqrt{1-tr\rho_3^2})-2+2d\\
=&&2\sqrt{d(d-1)}\sqrt{1-tr\rho_1^2}-\sqrt{d(d-1)}(\sqrt{1-tr\rho_2^2}+\sqrt{1-tr\rho_3^2})-2+2d\\
\geq&&2\sqrt{d(d-1)}\sqrt{\frac{d-1}{d}}-2+2d=0,
\end{eqnarray*}
where we have used $\sqrt{1-tr\rho_1^2}\geq0$ and $\sqrt{1-tr\rho_{k}^2}\leq 1-\frac{1}{d}$, $k=2$ or $3$ to obtain the last inequality above.

Thus we get
\be\sqrt{1-tr\rho_1^2}\geq \frac{1}{\sqrt{d(d-1)}}(\max \{M(\rho ), N(\rho )\}-\frac{1+2d}{3}).\ee
Similarly we obtain
 \be\sqrt{1-tr\rho_2^2}\geq \frac{1}{\sqrt{d(d-1)}}(\max \{M(\rho ), N(\rho )\}-\frac{1+2d}{3}).\ee
 \be\sqrt{1-tr\rho_3^2}\geq \frac{1}{\sqrt{d(d-1)}}(\max \{M(\rho ), N(\rho )\}-\frac{1+2d}{3}).\ee

Then according to the definition of GME concurrence, we derive
\be
C_{GME}(|\psi\ra)\geq \frac{1}{\sqrt{d(d-1)}}(\max \{M(\rho ), N(\rho )\}-\frac{1+2d}{3}).
\ee

Now we consider a mixed state $\rho\in H_1^d \otimes H_2^d \otimes H_3^d$ with the optimal ensemble decomposition
$\rho=\sum_{\alpha}p_{\alpha}|\psi_{\alpha}\ra\la\psi_{\alpha}|$ s.t. the GME concurrence attains its minimum.
One gets
\begin{eqnarray*}
C_{GME}(\rho)&=&\sum_{p_{\alpha},|\psi_{\alpha}\ra}p_{\alpha}C_{GME}(|\psi_{\alpha}\ra)\\
&\geq& \frac{1}{\sqrt{d(d-1)}}(\max \{M(\rho ), N(\rho )\}-\frac{1+2d}{3})\sum_{\alpha}p_{\alpha}\\
&=&\frac{1}{\sqrt{d(d-1)}}(\max \{M(\rho ), N(\rho )\}-\frac{1+2d}{3})
\end{eqnarray*}
which ends the proof of the theorem.\hfill \rule{1ex}{1ex}

\newpage
\bigskip
\noindent{\sf Acknowledgements}

\noindent This work is supported by the NSFC No.11775306, No.11675113, No.11701568; the Fundamental Research Funds for the Central Universities Grants No. 15CX05062A, No. 16CX02049A, and 17CX02033A; the Shandong Provincial Natural Science Foundation No.ZR2016AQ06; Qingdao applied basic research program No. 15-9-1-103-jch, and a project sponsored by SRF for ROCS, SEM.

\bigskip
\noindent{\sf Author contributions}

\noindent  M. Li and J. Wang wrote the main manuscript text. All
authors reviewed the manuscript.

\bigskip
\noindent{\sf Additional Information}

\noindent Competing Financial Interests: The authors declare no competing financial interests.


\begin{thebibliography}{99}


\bibitem{nielsen}Nielsen, M. A. \& Chuang, I. L. Quantum Computation and Quantum Information (Cambridge University Press, Cambridge,
England, 2000).

\bibitem{guhnerev}G$\ddot{u}$hne, O. \& T$\acute{o}$th, G. Entanglement detection. \textit{Phys. Rep.} \textbf{474}, 1-75 (2009).

\bibitem{mule1}Horodecki, R., Horodecki, P., Horodecki, M. \& Horodecki, K. Quantum entanglement. \textit{Rev. Mod. Phys.} \textbf{81}, 865 (2009).

\bibitem{mule2} Briegel H.J., Browne D.E., D$\ddot{u}$r W., Raussendorf R., \& Van den Nest M., Measurement-based quantum computation. \textit{Nat. Phys.} \textbf{5}, 19 (2009).


\bibitem{mule4} Gisin N., Ribordy G., Tittel W., \& Zbinden H. Quantum cryptography. \textit{Rev. Mod. Phy.} \textbf{74}, 145 (2002).

\bibitem{hillery} Hillery M., Bu$\check{z}$ek V. \& Berthiaume A. Quantum secret sharing. \textit{Phys. Rev. A} \textbf{59}, 1829 (1999).

\bibitem{srensen} Srensen A. S., \& Mlmer K. Entanglement and Extreme Spin Squeezing. \textit{Phys. Rev. Lett.} \textbf{86}, 4431 (2001).

\bibitem{toth} T$\acute{o}$th G. Multipartite entanglement and high-precision metrology. \textit{Phys. Rev. A} \textbf{85}, 022322 (2012).

\bibitem{rauss} Raussendorf R. \& Briegel H.J., A One-Way Quantum Computer. \textit{Phys. Rev. Lett} \textbf{86}, 5188 (2001).

\bibitem{mule3} Sen(De) A. \& Sen U. Quantum Advantage in Communication Networks. \textit{Phys. News} \textbf{40}, 17-32 (2010).

\bibitem{12} Huber M., Mintert F., Gabriel A., \&
Hiesmayr B.C. Detection of High-Dimensional Genuine Multipartite Entanglement of Mixed States. \textit{Phys. Rev. Lett.} \textbf{104}, 210501 (2010).

\bibitem{huber} Huber M. \& Sengupta R. Witnessing Genuine Multipartite Entanglement with Positive Maps. \textit{Phys. Rev. Lett.} \textbf{113}, 100501
(2014).

\bibitem{vicente3} Vicente, J. I. d. \& Huber, M. Multipartite entanglement detection from correlation tensors. \textit{Phys. Rev. A} \textbf{84}, 062306 (2011).

\bibitem{wu} Wu J.Y., Kampermann H., Bru${\ss}$ D., Kl$\ddot{o}$ckl C., \& Huber M. Determining lower bounds on a measure of multipartite entanglement from few local observables. \textit{Phys. Rev. A} \textbf{86}, 022319 (2012).

\bibitem{huber1} Huber M., Perarnau-Llobet M. \& Vicente J.I.
de. Entropy vector formalism and the structure of multidimensional entanglement in multipartite systems. \textit{Phys. Rev. A} \textbf{88}, 042328 (2013).

\bibitem{sperling} Sperling J. \& Vogel W. Multipartite Entanglement Witnesses. \textit{Phys. Rev. Lett.} \textbf{111}, 110503 (2013).



\bibitem{14} Eltschka C. \& Siewert J. Entanglement of Three-Qubit Greenberger-Horne-Zeilinger¨CSymmetric States. \textit{Phys. Rev. Lett.} \textbf{108}, 020502 (2012).


\bibitem{claude} Kl$\ddot{o}$ckl C., \& Huber, M. Characterizing multipartite entanglement without shared reference frames. \textit{Phys. Rev. A} \textbf{91}, 042339 (2015).

\bibitem{horo} M. Markiewicz, W. Laskowski, T. Paterek, and M. $\dot{Z}$ukowski
Phys. Rev. A 87, 034301 (2013).

\bibitem{ma1} Ma Z.H., Chen Z.H., Chen J.L.,
Spengler C., Gabriel A., \& Huber M. Measure of genuine multipartite entanglement with computable lower bounds. \textit{Phys. Rev. A} \textbf{83}, 062325(2011).

\bibitem{ma2} Chen Z.H., Ma Z.H., Chen J.L., \& Severini S. Improved lower bounds on genuine-multipartite-entanglement concurrence. \textit{Phys. Rev. A} \textbf{85}, 062320 (2012).

\bibitem{gaot1} Hong Y., Gao T., \& Yan F.L. Measure of multipartite entanglement with computable lower bounds. \textit{Phys. Rev. A} \textbf{86}, 062323 (2012).

\bibitem{gaot2} Gao T., Yan F.L., \& van Enk S.J. Permutationally Invariant Part of a Density Matrix and Nonseparability of N-Qubit States. \textit{Phys. Rev. Lett.} \textbf{112}, 180501
(2014).

\bibitem{bellgme} Bancal J.D., Gisin N., Liang Y.C., \& Pironio S. Device-Independent Witnesses of Genuine Multipartite Entanglement. \textit{Phys. Rev. Lett.} \textbf{106}, 250404 (2011).

\bibitem{8} Eltschka C., \& Siewert J. Quantifying entanglement resources. \textit{J. Phys. A: Math.Theor.} \textbf{47}, 424005 (2014).


\bibitem{jungnitsch} Jungnitsch B., Moroder T., \& G$\ddot{u}$hne O. Taming Multiparticle Entanglement. \textit{Phys. Rev. Lett.} \textbf{106}, 190502 (2011).

\bibitem{10} Lancien C., G$\ddot{u}$hne O., Sengupta R., \& Huber M. Relaxations of separability in multipartite systems:
Semidefinite programs, witnesses and volumes. \textit{J. Phys. A: Math. Theor.} \textbf{48}, 505302 (2015).

\bibitem{ppt} Peres A. Separability Criterion for Density Matrices. \textit{Phys. Rev. Lett.} \textbf{77}, 1413 (1996).

\bibitem{san98} Sanpera A., Tarrach R., \& Vidal G. Local description of quantum inseparability. \textit{Phys. Rev. A} \textbf{58}, 826 (1998).

\bibitem{rudolph} Rudolph O. On the cross norm criterion for separability. \textit{J. Phys. A: Math. Theor.} \textbf{36}, 5825 (2003).


\bibitem{ChenQIC03} Chen, K. \& Wu, L. A. A matrix realignment method for recognizing entanglement. \textit{Quantum Inf. Comput.} \textbf{3}, 193¨C202 (2003).

\bibitem{chenprl} Chen K., Albeverio S., \& Fei S.-M. Concurrence of Arbitrary Dimensional Bipartite Quantum States. \textit{Phys. Rev. Lett.} \textbf{95}, 040504 (2005).


\bibitem{zhangtgpra}  Zhang T.G., Zhao M.J., Li M., Fei S.M. \& Li-Jost X.Q. Criterion of Local Unitary Equivalence for Multipartite States. \textit{Phys. Rev. A} \textbf{88}, 042304 (2013).

\end{thebibliography}
\end{document}